\documentclass[useAMS,usenatbib]{mn2e}
\usepackage{graphicx}

 \title[The radial distribution of magnetic helicity in the solar
convective zone]{The radial distribution of magnetic helicity in
the solar convective zone: observations and dynamo theory}

 \author[H.\, Zhang, D.\, Sokoloff, I.\, Rogachevskii, D.\, Moss,
 V. \, Lamburt, K.\, Kuzanyan and N.\, Kleeorin]
 {H.\, Zhang, $^{1}$ D.\, Sokoloff, $^{2}$
 I.\, Rogachevskii, $^{3}$ D.\, Moss, $^{4}$\thanks{E-mail: moss@maths.man.ac.uk}
 V. \, Lamburt, $^{2}$
 \newauthor
 K.\, Kuzanyan $^{5}$ and N.\, Kleeorin $^{3}$\\
 $^{1}$ National Astronomical
   Observatories, Chinese Academy of Sciences, Beijing 100012, China\\
 $^{2}$ Department of Physics, Moscow
   State University, Moscow 119992, Russia \\
 $^{3}$ Department of Mechanical Engineering,
        Ben-Gurion University of Negev, POB 653, 84105 Beer-Sheva, Israel\\
 $^{4}$ School of Mathematics, University of Manchester,
        Oxford Rd, Manchester M13 9PL, UK \\
$^{5}$   IZMIRAN, Troitsk, Moscow Region 142190, Russia}

\begin{document}

 \date{Accepted . Received ; in original form}

\pagerange{\pageref{firstpage}--\pageref{lastpage}} \pubyear{2005}

\maketitle

\label{firstpage}

\begin{abstract}
   We continue our attempt to connect observational data
   on current helicity in solar active regions with solar
   dynamo models.
   In addition to our previous results about
   temporal and latitudinal distributions of current helicity
   (Kleeorin et al. 2003),
   we argue that some information concerning the radial profile of
   the current helicity averaged over time and latitude can be
   extracted from the available observations. The main feature of
   this distribution can be presented as follows. Both shallow and
   deep active regions demonstrate a clear dominance of one
   sign of current helicity in a given hemisphere during the whole
   cycle. Broadly speaking, current helicity has opposite polarities in the
   Northern and Southern hemispheres, although there are some active regions
   that violate
   this polarity rule. The relative number of active regions
   violating the polarity rule is significantly higher for deeper
   active regions. A separation of active regions into `shallow',
   `middle' and `deep' is made by comparing their rotation
   rate and the helioseismic rotation law. We use a
   version of Parker's dynamo model in two spatial dimensions,
   that employs a nonlinearity based on
   magnetic helicity conservation arguments. The predictions of this model
   about the radial distribution of solar current helicity
   appear to be in remarkable agreement with the available
   observational data; in particular the relative volume occupied by the current
   helicity of "wrong" sign grows significantly with the depth.
\end{abstract}


\section{Introduction}

The solar 22-year activity cycle is thought to be a manifestation
of dynamo action somewhere inside the solar convective zone or
even in the overshoot layer. The solar differential rotation acts
as a driver of the solar dynamo, generating a toroidal magnetic
field from an existing poloidal magnetic field. The other dynamo
driver, required to transform toroidal magnetic field into
poloidal and so to close the chain of self-excitation, is thought
to be what is commonly known as the $\alpha$-effect, i.e. a
specific feature of convective flows in a rotating body. It was
E.~Parker who suggested as early as 1955 that cyclonic motions in
the solar convective zone produce a mean (large-scale) poloidal
magnetic field from a mean toroidal magnetic field. Ten years
later, Steenbeck, Krause \& R\"adler developed a theory of this
process, calling it the $\alpha$-effect (see Krause \& R\"adler
1980). A physical feature of the $\alpha$-effect in the form
discussed at this stage is that the action of the Coriolis force
on the convective vortices results in a domination of right-handed
vortices in the Northern solar hemisphere and, correspondingly,
left-handed vortices in the Southern. A non-vanishing difference
between vortices with right and left helicities in a given
hemisphere provides the required conversion of toroidal magnetic
field to poloidal.

Parker (1955) demonstrated that the scheme briefly discussed above
leads to the self-excitation of a wave of magnetic field (the
so-called dynamo wave). A suitable choice of the differential
rotation shear and mean helicity of solar convection in, say, the
Northern hemisphere leads to a dynamo wave whose shape mimics
remarkably that of the solar butterfly diagram. The simplest order
of magnitude estimates for the dynamo governing parameters results
in an estimate for the cycle length which is about 10 times
shorter then the real solar cycle -- this seems reasonable
for this obviously oversimplified model.

Until now the above scheme for the solar dynamo, known as the
Parker migratory dynamo, has remained the basis of most dynamo
models for solar and stellar dynamo activity. Of course, present
day solar dynamo models include achievements of helioseismology,
effects of meridional circulations and various other features of
solar MHD. As a result, these dynamo models are much richer and,
in principle at least, closer to the real Sun then the simple
Parker model. Nevertheless, although these more sophisticated
models can reproduce many specific details, some points remain as
obscure now as in 1955 (for recent reviews, see Ossendrijver 2003;
Brandenburg \& Subramanian 2004).

Note that the $\alpha$-effect remained for several decades a
theoretical concept only. It is deeply associated with
the helicity of rotating turbulence and arises from
averaging Maxwell's equations over the ensemble of rotating
vortices. For a long time, there was no evidence available to
support the $\alpha$-effect from either astronomical observations
or from laboratory MHD experiments. Obviously, such a situation
makes the basis of solar dynamo theory rather unsatisfactory and
even shaky.

In the last decade, some basic progress here has been made and the
first observational data of physical quantities associated with
the $\alpha$-effect are now available. The fundamental point is
that the $\alpha$-effect includes two contributions (Pouquet et
al. 1976),  an hydrodynamical contribution as discussed above
($\alpha^v$) associated with helicity of convective vortices, and
also a contribution from the helicity of the magnetic field itself
($\alpha^m$). The hydrodynamic helicity is determined by a
correlation between the convective velocity $\vec{u}$ and its
vorticity, i.e. $ \langle\vec{u} \cdot (\vec{\nabla} \times
\vec{u})\rangle$, and so its observational determination requires
knowledge of all three components of velocity while the Doppler
effect gives a line-of-sight velocity component only. The magnetic
part of the $\alpha$-effect, $\alpha^m$, can be related to what
has become known as the current helicity, proportional to $
\langle\vec{b} \cdot (\vec{\nabla} \times \vec{b}) \rangle$, where
$\vec{b}$ is the small-scale magnetic field. Because Zeeman
splitting provides information concerning all 3 components of
$\vec{b}$, $\,\,\,\alpha^m$ appears to be more accessible for
observational determination than $\alpha^v$ (Seehafer 1990). As a
result, the first observations to be made relate to the current
helicity in active regions on the solar surface (Pevtsov et al.
1994, 1995; Zhang \& Bao 1998, 1999; Longcope et al. 1998).

Such observational findings about the current helicity on the
solar surface can be related to theoretical results in dynamo
theory, where the concept of the magnetic part of $\alpha$-effect
has been developed into a theory of dynamo saturation through
$\alpha^m$. Kleeorin \& Ruzmaikin (1982) and Kleeorin \&
Rogachevskii (1999) suggested a governing equation for $\alpha^m$
which describes the time evolution of the $\alpha$-effect.
Together with the mean-field dynamo equations, this equation has
solutions in the form of a propagating dynamo wave which amplitude
is steady in time (Kleeorin et al. 1995; Covas et al. 1998;
Blackman \& Brandenburg 2002).

Kleeorin et al. (2003) discussed a link between the observational
and theoretical findings outlined above. They concluded that the
accumulated observational knowledge is sufficient to follow the
temporal evolution during one solar cycle of current helicity
averaged over a given hemisphere or the latitudinal distribution
of current helicity averaged over one solar cycle. Existing
ideas concerning the nonlinear solar dynamo saturated by the
magnetic part of the $\alpha$-effect provide a theoretical
prediction of the corresponding quantities. These demonstrate a
general agreement with observations and provide a possibility of
fitting the governing parameters of the solar dynamo by
observational data.

Here we present an extension of the approach of the paper of
Kleeorin et al. (2003). First of all, we discuss the extent to
which the solar helicity data can be used to understand the radial
dependence of solar magnetic helicity and the corresponding dynamo
activity. An initial step in this direction was made by Kuzanyan
et al. (2003) who separated the database of active regions for
which the helicity data are available into subsets corresponding
to shallow, middle and deep active regions, according to their
rotation rate. Of course, a substantial part of the data cannot be
so classified. After averaging magnetic helicity data over the
subsets, we obtain quantities which can be compared with the
theoretical data averaged over the three radial ranges.

This approach is to some extent similar to the studies of the
solar rotation curve using the sunspot data associated with
various types of sunspots. Following the remarkable advances in
helioseismology such reconstructions now look rather archaic (this is
why we refer here only to the single paper, Collin et al. (1995),
in which one of the authors participated). However, at this
preliminary stage of solar helicity studies, a similar approach
appears to be reasonable.

The other topic to be addressed here is a plausible improvement of
the governing equation for the magnetic helicity. The point is
that the dynamo saturation by a magnetic contribution to the
$\alpha$-effect is necessarily combined with a modification of the
turbulent diffusivity and other transport coefficients. For the
sake of simplicity, these effects were ignored in our first paper
(Kleeorin et al. 2003). Now we restore these terms in order to
investigate their possible contribution, and to produce a more
fully self-consistent model.

Obviously, our model is still simplified and does not include many
important features of solar activity. In particular we do not
address the problem of the storage of magnetic fields and the
formation of flux tubes in the overshoot layer near the bottom of
the convective zone (see, e.g., Spiegel \& Weiss 1980; Tobias et
al. 2001; Tobias \& Hughes 2004; Brandenburg 2005, and references
therein).

\section{Data on current helicity obtained at the Huairou Solar Observatory
\label{observs} Station}

Our research is based on the data on current helicity accumulated
during 10 successive years (1988-97) of observations at the
Huairou Solar Observatory Station of the National Astronomical
Observatories of China (Bao and Zhang 1998),
which were further processed by
Zhang et al. (2002). The description of the observational
procedure and the basic ideas of data processing can be found in
Kleeorin et al. (2003) and references therein. The total available
sampling that we used contains data of 410 active regions.

Following Kuzanyan et al. (2003) we divide the active region into
4 groups, i.e. shallow, middle and deep active regions, as well as
a group for which the depth cannot be estimated satisfactorily
(Table~1). The separation of the active regions into three groups
is based on the result of  helioseismology (Schou et al. 1998)
that the angular rotation rate growth monotonically with radius at
least for the domain between fractional radii 0.65 and 0.95 and
latitudes below 30-35$^\circ$ (for details see Kuzanyan et al.
2003).

The Solar Geophysical Data records, which can be obtained from the
NOAA (USAF-MWL) database, provide us with several tens of
longitudinal locations (in terms of the Carrington coordinate
system) for each active region under investigation, for several
consequent days. Therefore, we attempt to calculate partial, or
``individual'', angular rotation rates with respect to  the
Carrington rotation. For some active regions we can find a certain
trend in the evolution of their Carrington coordinates with time.
From the complete sampling of the data, which contain 410 active
regions, we select subsamples for which this trend in Carrington
longitude {\it versus} time has significant correlation. We
determined the subsamples for which the correlation coefficient
$\sigma$ is greater than 0.5 and 0.6 respectively. These samples
contain 178 and 134 active regions (or $43\%$ and $33\%$ of the
available data), respectively.

Given an ``individual'' angular rotation rate for each active
region we can identify them with certain {\it effective} depths.
Using a particular analytical approximation of the solar rotation
curve (see Kuzanyan et al. 2003), the active regions with known
individual angular rotation were separated into three groups. The
individual rotation rates in the first group fall into the range
covered by the analytical approximation for the radial range $r
\le 0.76$, for the second group the bound is $0.76 \le r \le
0.84$, and $r>0.84$ for the third group. These groups were labelled
as deep, middle and shallow. Notice, that the internal rotation of
the solar convective zone above approximately fractional
radius 0.94 is slower than in the zone below, and so we
disregarded this sub-surface layer. We
stress that the above bounds were chosen rather arbitrarily and
the details of trends in the current helicity properties with
respect to depth can hardly be considered quantitatively. The
active regions appear to be distributed between the upper and
lower layers approximately equally, while very few occur
within the middle layer (Kuzanyan et al. 2003). We will consider
separately the upper and lower layers and compare the results of
statistical analysis of the data in each of them.

Because the current helicity is expected to be of opposite sign in
Northern and Southern hemispheres, we subdivide these groups
between the two hemispheres and average the data in each group
over all latitudes as well as cycle phases. The result of
averaging $H_c$ is given in Table~1 for active regions with
identified depth. Here $d$ is a depth identifier, with "s" meaning
shallow, "m" middle and "d" deep active regions. Because the
number of active areas of intermediate depth appears to be quite low,
and insufficient to estimate the sign of helicity, we combine
quite arbitrarily the data for the middle and deep active regions
into a single group, i.e. "d+m". $N$ is the number of active
regions included in each group. For Table~1, we use the threshold
$\sigma=0.5$. To demonstrate the stability of the selection
procedure to the threshold value, we give in Table~2 similar
results for the threshold value $\sigma=0.6$.

\begin{table}
\label{tab1}
\begin{tabular}{|l|c|c|c|c|}
\multicolumn{5}{c}{Table 1}\\
\multicolumn{5}{c}{Current helicity $H_c$ for active regions}\\
\multicolumn{5}{c}{binned by
depth, threshold $\sigma=0.5$ Here and} \\
\multicolumn{5}{c}{below $H_c$ is measured in in units of
$10^{-3} {\rm G}^2 {\rm m}^{-1}$}\\
\hline
$d$ & $N$ & $N^*$ & $H_c$ & $N^*/N$\\
\hline
\multicolumn{5}{|c|}{North}\\
\hline
s & 47 & 1 & $-0.6 \pm 0.2$ & $0.02\pm 0.04$ \\
\hline
m & 5 & 1 & $-0.2 \pm 0.7$ & $0.20 \pm 0.35$\\
\hline
d & 34 & 8 & $-1.0 \pm 0.7$ & $0.24 \pm 0.14$\\
\hline
d+m & 39 & 9 & $-0.9 \pm 0.6$ & $0.23\pm 0.13$\\
\hline
\multicolumn{5}{|c|}{South}\\
\hline
s & 41 & 5 & $0.5 \pm 0.6$ & $0.12 \pm 0.10$\\
\hline
m & 6 & 2 & $0.3 \pm 1.5$ & $0.33\pm 0.38$ \\
\hline
d & 38 & 11 & $0.6 \pm 0.4$ & $0.29\pm 0.14$\\
\hline
d+m & 44 & 13 & $0.6 \pm 0.4$ & $0.3 \pm 0.13$ \\
\hline
\end{tabular}
\end{table}

\begin{table}
\label{tab2}
\begin{tabular}{|l|c|c|c|c|}
\multicolumn{5}{c}{Table 2}\\
\multicolumn{5}{c}{Current helicity for active regions}\\
\multicolumn{5}{c}{binned by
depth, threshold $\sigma=0.6$}\\
\multicolumn{5}{c}{}\\
\hline \multicolumn{5}{|c|}{North}\\
\hline
$d$ & $N$ & $N^*$ & $H_c$ & $N^*/N$\\
\hline
s & 33 & 1 & $-0.6 \pm 0.3$ & $0.03\pm 0.06$ \\
\hline
m & 2 & 1 & $-0.3 \pm 9.4$ & $0.5 \pm 0.69$ \\
\hline
d & 28 & 7 & $-1.0 \pm 0.8$ & $0.25 \pm 0.16$\\
\hline
d+m & 30 & 8 & $-1.0 \pm 0.8$ & $0.27 \pm 0.16$ \\
\hline \multicolumn{5}{|c|}{South}\\
\hline
s & 33 & 4 & $0.6 \pm 0.7$ & $0.12\pm 0.11$\\
\hline
m & 3 & 2 & $-0.2 \pm 4.8$ & $0.7\pm 0.53$\\
\hline
d & 29 & 9 & $0.4 \pm 0.4$ & $0.31\pm 0.17$\\
\hline
d+m & 32 & 11 & $0.4 \pm 0.4$ & $0.34\pm 0.16$\\
\hline
\end{tabular}
\end{table}

In agreement with theoretical expectations, the data for $H_c$ are
remarkably antisymmetric in respect to the solar equator. Note
that the same kind of antisymmetry was recognized in the averaging
over latitude or time undertaken in Kleeorin et al. (2003). We
note however that there are a significant number of active regions
that violate this polarity law. The number of such active regions
are given in Tables~1 and 2 as $N^*$.

We present in Table~3 the averaged values of the helicities of for
all 410 active regions for which the observations of helicity are
available. These active regions follow the same polarity rule as
the active regions with known depth, and again some active regions
violate this rule. Their number is given as $N^*$.

\begin{table}
\label{tab3}
\begin{tabular}{|l|c|c|c|c|}
\multicolumn{5}{c}{Table 3}\\
\multicolumn{5}{c}{Current helicity for all 410 active regions}\\
\multicolumn{5}{c}{}\\
\hline
hemisphere & $N$ & $N^*$ & $H_c$ & $N^*/N$\\
\hline
North & 193 & 30 & $-0.8 \pm 0.2$ & $0.16\pm 0.05$\\
\hline
South & 217 & 47 & $0.6 \pm 0.2$ & $0.22\pm 0.05$\\
\hline
\end{tabular}
\end{table}

The number of active regions with current helicity that violate
the polarity rule can be calculated for both hemispheres
(Table~4). Note that it is not appropriate to average the current
helicity over both hemispheres because the data in the Northern
and Southern hemisphere cancel. We conclude from Table~4 that the
deep (and middle) active regions contain several times more cases
of parity rule violations than the shallow active regions, and
even slightly more then the active regions without definite
estimation of depth.

We were unable to recognize any clear trend in the number of
active regions violating the polarity rule selected according to
latitude or the cycle phase. However we present the relevant data
below (Tables 5 and 6).

\begin{table}
\label{tab4}
\begin{tabular}{|l|c|c|c|}
\multicolumn{4}{c}{Table 4}\\
\multicolumn{4}{c}{Number of active region with current}\\
\multicolumn{4}{c}{helicity violating the polarity rule,}\\
\multicolumn{4}{c}{binned by depth, threshold $\sigma=0.5$.}\\
\multicolumn{4}{c}{}\\
\hline
depth & $N$ & $N^*$ & $N^*/N$\\
\hline
s & 88 & 6 & $0.07\pm 0.05$\\
\hline
d+m & 83 & 22 & $0.27\pm 0.09$\\
\hline
\end{tabular}
\end{table}

\begin{table}
\label{tab5}
\begin{tabular}{|l|c|c|c|}
\multicolumn{4}{c}{Table 5}\\
\multicolumn{4}{c}{Number of active region with current}\\
\multicolumn{4}{c}{helicity violating the polarity rule}\\
\multicolumn{4}{c}{ordered by date, threshold $\sigma=0.5$.}\\
\hline
years & $N$ & $N^*$ & $N^*/N$\\
\hline
1988-89 & 87 & 23 & $0.26 \pm 0.09$\\
\hline
1990-91 & 126 & 20 & $0.16 \pm 0.06$\\
\hline
1992-93 & 121 & 18 & $0.15 \pm 0.06$\\
\hline
1994-96 & 69 & 13 & $0.18 \pm 0.09$\\
\hline
\end{tabular}
\end{table}

\begin{table}
\label{tab6}
\begin{tabular}{|l|c|c|c|}
\multicolumn{4}{c}{Table 6}\\
\multicolumn{4}{c}{Number of active region with current}\\
\multicolumn{4}{c}{helicity violating the polarity rule,}\\
\multicolumn{4}{c}{ordered by latitude $\Theta$, threshold $\sigma=0.5$.}\\
\hline
latitude (degrees) & $N$ & $N^*$ & $N^*/N$\\
\hline
$24\le \Theta \le 32$ & 18 & 4 & $0.22 \pm 0.19$\\
\hline
$16 \le \Theta \le 24$ & 53 & 10 & $0.19 \pm 0.11$\\
\hline
$12 \le \Theta \le 16$ & 36 & 5 & $0.14 \pm 0.11$\\
\hline
$8 \le \Theta \le 12$ & 48 & 8 & $0.17 \pm 0.11$\\
\hline
$-8 \le \Theta \le 8$ & 65 & 6 & $0.08 \pm 0.06$\\
\hline
$-12 \le \Theta \le -8$ & 58 & 12 & $0.21 \pm 0.10$\\
\hline
$-16 \le \Theta \le -12$ & 46 & 8 & $0.17 \pm 0.11$\\
\hline
$-24 \le \Theta \le -16$ & 67 & 19 & $0.28 \pm 0.11$\\
\hline
$-32 \le \Theta \le -24$ & 12 & 3 & $0.25 \pm 0.25$\\
\hline
\end{tabular}
\end{table}

\section{The dynamo model}

We use here a dynamo model which is basically an extension of the
simplified model of Kleeorin et al. (2003). In particular, the
present model includes an explicit radial coordinate and takes
into account the curvature of the convective shell, and also
quenching of turbulent magnetic diffusivity. We start from the
general mean-field dynamo equations (see e.g. Moffatt 1978, Krause
\& R\"adler 1980). Using spherical coordinates $r, \theta, \phi$
we describe an axisymmetric magnetic field by the azimuthal
component of magnetic field $B$, and the component $A$ of
the magnetic potential corresponding to the
poloidal field. Following Parker (1955) we consider dynamo
action in a convective shell. However we retain a radial
dependence of $A$ and $B$ in the dynamo equations and we do not
neglect the curvature of the shell. The equations for
$\tilde A = r \sin \theta A$ and  $\tilde B = r \sin \theta \, B$
read

\begin{eqnarray}
&& {\partial \tilde  A \over \partial t} + {V^A_\theta \over r}
   {{\partial \tilde A} \over {\partial \theta}} + V^A_r
   {{\partial \tilde A} \over {\partial r}} = C_\alpha \, \alpha \,
   \tilde B + \eta_{_{A}} \biggl[{{\partial ^2 \tilde A }
   \over {\partial r^2}}
   \nonumber \\
&& \; \; \; \; + {\sin \theta \over r^2} {\partial
   \over \partial \theta} \biggl({1 \over \sin \theta} {\partial
   \tilde A \over \partial \theta}\biggr) \biggr] \;,
\label{L1} \\
&& {{\partial \tilde B} \over {\partial t}} +  {\sin \theta \over
   r} {\partial \over \partial \theta} \biggl({V^B_\theta \tilde B
   \over \sin \theta}\biggr) + {{\partial (V^B_r \tilde B)} \over
   {\partial r}} = \sin \theta \, \biggl( G_r {{\partial } \over
   {\partial \theta}}
   \nonumber \\
&&  \; \; \; \; - G_\theta {{\partial } \over {\partial r}}
   \biggr) \tilde A + {\sin \theta \over r^2} {\partial \over
   \partial \theta} \biggl({\eta_B \over \sin \theta} {\partial
   \tilde B \over \partial \theta}\biggr)
   + {\partial \over {\partial r}} \biggl( \eta_{_{B}}
{{\partial \tilde B} \over {\partial r}} \biggr) \;,
\nonumber \\
\label{L2}
\end{eqnarray}
where

\begin{eqnarray*}
G_r = {{\partial \Omega} \over {\partial r}} \;, \quad G_\theta =
{{\partial \Omega} \over {\partial \theta}} \; .
\end{eqnarray*}
Here we measure lengths in units of the solar radius $R_\odot$ and
time in units of a diffusion time based on the solar radius and
the turbulent magnetic diffusivity $\eta_{_{T0}}$. When estimating
this timescale we use the `basic' (assumed uniform) value  of the
turbulent magnetic diffusivity, unmodified by the magnetic field.

We consider the fractional radial range $0.64 <r<1$, where $r =
0.64$ corresponds to the bottom of the convective zone and $r=1$
corresponds to the solar surface. The `convection zone' proper can
be thought of as occupying $0.7\le r\le 1.0$, with $0.64\le r \le
0.7$ being a tachocline/overshoot region. The rotation law
includes radial shear (proportional to $G_r$) and a latitudinal
dependence (proportional to $G_\theta$).

At the surface $r=1$ we use vacuum boundary conditions on the
field, i.e. $B=0$ and the poloidal field fits smoothly onto a
potential external field. At the lower boundary, $r=r_0=0.64$,
$B=B_r=0$. At both $r=r_0$ and $r=1$, $\partial \chi^c/\partial
r=0$, where $\chi^c$ is the current helicity (see
Eq.~(\ref{helic})).

Of course these equations, although more elaborate than those
often used to study the solar cycle, are still oversimplified.
However they appear adequate to reproduce the basic qualitative
features of solar (and stellar) activity. Taking into account the
exploratory nature of the approach, we use the simplest profiles
of dynamo generators compatible with symmetry requirements and
with producing a magnetic butterfly diagram that is concentrated
towards low latitudes (see also R\"udiger \& Brandenburg 1995;
Moss \& Brooke 2000). Thus the unquenched hydrodynamical part of
the $\alpha$-effect,  $\alpha^v(B=0) = \chi^v = \sin^2 \theta \,
\cos \theta$ and $C_\alpha < 0$ (this determines  the sign value
of the hydrodynamic $\alpha$ effect, see below in Sect.~4). The
points $\theta = 0$ and $\theta = 180^\circ$ correspond to the
North and South poles respectively. See Kleeorin et al. (2003) for
further discussion of this approach.

As a new feature of Eqs.~(\ref{L1}) and (\ref{L2}), compared with
the dynamo model exploited by Kleeorin et al. (2003), we retain
here the possibility of including a contribution from the dynamo
generated magnetic field in the turbulent diffusion coefficients
($\eta_{_{A}}$ and $\eta_{_{B}}$), and the meridional circulation
($V^A_\theta, V^A_r, V^B_\theta$ and $V^B_r$). However we do not
consider fully here the role of meridional circulation.

The magnetic field is measured in units of the equipartition field
$ B_{\rm eq} = u \sqrt{4 \pi \rho_\ast}$,  and the vector
potential of the poloidal field $A$ is measured in units of
$R_\odot B_{\rm eq}$. The density $\rho$ is normalized with
respect to its value $\rho_\ast$ at the bottom of the convective
zone, and the basic scales of the turbulent motions $l$ and
turbulent velocity $u$ at the scale $l$ are measured in units of
their maximum values through the convective zone. Because
turbulent diffusivity and $\alpha$-effect depend on the magnetic
field, we use their initial values in the limit of very small mean
magnetic field to obtain the dimensionless form of the equations.
To emphasize this, we do not introduce the dynamo number in an
explicit form here however use it below when convenient.

\section{The nonlinearities}

We present below a model for the nonlinear dynamo saturation. The
model is based as far as possible on first principles, and is
similar to that used in the derivation of the equations of
mean-field electrodynamics by Krause \& R\"adler (1980). As an
important technical point, we used a quasi-Lagrangian approach
in the framework of Wiener path integrals to derive the dynamical
equation for the evolution of the magnetic helicity including
magnetic helicity flux (see Kleeorin \& Rogachevskii 1999). We
also used the spectral $\tau$-approximation (Orszag's third-order
closure procedure) to determine the nonlinear mean electromotive
force (see Rogachevskii \& Kleeorin 2000, 2004). Here we note some
important features of the model only.

A key assumption of the model under discussion is the concept of
the locally isotropic and weakly inhomogeneous nature of the
background MHD turbulence (with a zero mean magnetic field).
Because we include large-scale phenomena such as helicity
advection, the accuracy of the approximation is limited. In
particular, a completely rigorous evaluation of the turbulent
diffusion of magnetic helicity is beyond the scope of our model
and we allow this quantity to be transported by the turbulent
diffusion in the same way as a scalar admixture, i.e. the
turbulent diffusion coefficient is determined by the velocity
field correlation tensor. In contrast, the nonlinear coefficients
of the large-scale magnetic field defining the nonlinear mean
electromotive force are determined by the cross-helicity of
magnetic ($b_i$) and velocity ($u_i$) fields, i.e.  by $\langle b_i
u_i \rangle$. The different scalings for these quantities
presented below are connected with this fact.

Note that a deeper investigation of the turbulent diffusion of
magnetic helicity, as well as of the non-diffusive fluxes of
magnetic helicity, looks possible in principle. It would require
at least the application of Orszag's fourth-order closure
procedure to derive the magnetic helicity fluxes. However, this
generalization would require a much more extended calculation than
required to obtain the model considered here. As a substantial
body of calculations already have been necessary, it seems very
reasonable to clarify the astrophysical consequences of the model
now available, before attempting to move on further.

We stress again that the model analyzed is derived, as far as possible,
from first principles. The scope of the model is however
obviously limited and does not include all possible physical
mechanisms which could in principle contribute to dynamo
saturation. In particular, we do not include the buoyancy of the
magnetic field. Some other limitations are mentioned below.
Bearing in mind the natural limitations of the model, we introduce
several numerical coefficients $C_1, C_2, C_3$ multiplying the
magnetic helicity fluxes, which we consider to be free parameters
of order unity (see Eq.~(\ref{curhel}) below).

\subsection{The $\alpha$-effect}

The key idea of the dynamo saturation scenario exploited below (as
well as by Kleeorin et al. 2003) is the splitting of the total
$\alpha$-effect into its hydrodynamic and magnetic parts,
$\alpha^v$ and $\alpha^m$ respectively. The
calculation of the magnetic part of the $\alpha$-effect is based
on the idea of magnetic helicity conservation and the link between
current and magnetic helicities, and gives (see Kleeorin et al.
2000, 2003)

\begin{equation}
\alpha = \alpha^v + \alpha^m = \chi^v \phi_v
+ {{ \phi_m} \over {\rho(z)}}\chi^c \;.
\label{helic}
\end{equation}
Here $\chi^v$ and $\chi^c$ are proportional to the hydrodynamic
and current helicities respectively and $\phi_v$ and $\phi_m$ are
quenching functions. The analytical form of the quenching
functions $\phi_{v}(B)$ and $\phi_{m}(B)$ is given in Appendix A.
In contrast to Kleeorin et al. (2003), we consider here the radial
helicity profiles in an explicit form and so we keep in
Eq.~(\ref{helic}) the radial profile of density $\rho (z)$
normalized by the density $\rho_\ast$ at the bottom of the
convective zone. This factor appears as $\chi^c = (\tau/12 \pi
\rho_\ast) \langle \vec{b} \cdot (\vec{\nabla} \times \vec{b})
\rangle$ (for details, see Kleeorin et al. 2003). Based on Baker
\& Temesvary (1966) and Spruit (1974), we choose for $\rho(z)$ the
analytical approximation

\begin{equation}
\rho(z) = \exp [- a \tan (0.45 \pi \, z)] \;, \label{rho}
\end{equation}
where $z=1 - \mu (1-r) $ and $\mu = (1 - R_0/R_\odot)^{-1}$. Here
$a \approx 0.3$ corresponds to a tenfold change of the density in
the solar convective zone, $a \approx 1$ by a factor of about $10^3$, etc.
However in the majority of our investigations we took $\rho={\rm
const.}$, but we did also consider cases with $a=0.3$.

The equation for $\tilde \chi^c = r^2 \sin^2 \theta \, \chi^c$
is
\begin{eqnarray}
&& {{\partial \tilde \chi^c} \over {\partial t}} + {{\tilde\chi^c}
\over T} = \left({{2R_\odot} \over l}\right)^2 \biggl\{ {1 \over
C_\alpha} \biggl[{\eta_{_{B}} \over r^2} \, {{\partial \tilde A}
\over {\partial \theta}} {{\partial \tilde B} \over {\partial
\theta}} + \eta_{_{B}} \, {{\partial \tilde A} \over {\partial r}}
{{\partial \tilde B} \over {\partial r}}
\nonumber\\
&& \; \; - \eta_{_{A}} \, \tilde B \, {\sin \theta \over r^2}
{\partial \over \partial \theta} \biggl({1 \over \sin \theta}
{\partial \tilde A \over \partial \theta}\biggr) - \eta_{_{A}} \,
\tilde B {{\partial ^2 \tilde A} \over {\partial r^2}}
\nonumber\\
&& \; \; + (V^A_r - V^B_r) \, \tilde B {{\partial \tilde A} \over
{\partial r}} + (V^A_\theta - V^B_\theta) {\tilde B \over r}
{{\partial \tilde A} \over {\partial \theta}} \biggr] - \alpha
\tilde B^2 \biggr\}
\nonumber\\
&& \; \; - {{\partial \tilde {\cal F}_r} \over {\partial r}} -
{\sin \theta \over r} {{\partial} \over {\partial \theta}}
\left({\tilde {\cal F}_\theta \over  \sin \theta} \right)  \;,
\label{curhel}
\end{eqnarray}
where $\vec{\tilde {\cal F}} = r^2 \sin^2 \theta  \vec{\cal F}$,
and the flux of the magnetic helicity is chosen in the form
\begin{eqnarray}
\vec{\cal F} &=& \eta_{_{A}}(B) \, B^2 \, \{C_1 \, \vec{\nabla}
[\chi^v \, \phi_v(B)] + C_2 \, \chi^v \, \phi_v(B) \,
\vec{\Lambda}_{\rho} \}
\nonumber\\
&& - C_3 \, \kappa \, \vec{\nabla} \, \chi^c  \;, \label{flux}
\end{eqnarray}
with $\vec{\Lambda}_\rho = - \vec{\nabla} \rho / \rho$. Here
$R_\odot/l$ is the ratio of the solar radius to the basic scale of
solar convection, $ T = (1/3) \, {\rm Rm} \, (l/R_\odot)^2 $ is
the dimensionless relaxation time of the magnetic helicity, ${\rm
Rm} = lu / \eta_0 $ is the magnetic Reynolds number, with
$\eta_0$ the `basic' magnetic diffusion due to the electrical
conductivity of the fluid. Equation~(\ref{curhel}) is a
generalization of Eq.~(A.3) of Kleeorin et al. (2003) to the case
considered here. The fluxes of magnetic helicity~(\ref{flux})
were derived using Eqs.~(9) and ~(13) of Kleeorin \& Rogachevskii
(1999). Equation~(\ref{flux}) is in agreement with the results
of Vishniac \& Cho (2001) and Subramanian \& Brandenburg (2004).

Let us estimate the values of the governing parameters for
different depths of the convective zone. We stress that all
physical ingredients of the model vary strongly with the depth
$h_\ast$ below the solar surface. We use mainly estimates of
governing parameters taken from models of the solar convective
zone, e.g. Spruit (1974) and Baker \& Temesvary (1966) -- more
modern treatments make little difference to these estimates. In
the upper part of the convective zone, say at depth $ h_\ast \sim
2 \times 10^7$ cm (measured from the top), the parameters are $
{\rm Rm} \sim 10^5 ,$ $\, u \sim 9.4 \times 10^4 $ cm s$^{-1}$, $
l \sim 2.6 \times 10^7$ cm, $ \, \rho \sim 4.5 \times 10^{-7}$ g
cm$^{-3} ,$ $\, \eta_{_{T}} \,(\rm {the\,\, turbulent\,\,
diffusivity}) \sim 0.8 \times 10^{12} $ cm$^2$ s$^{-1}$; the
equipartition mean magnetic field is $B_{\rm eq} \sim 220 $ G and
$ T \sim 5 \times 10^{-3}$. At depth $ h_\ast \sim 10^9$ cm these
values are $ {\rm Rm} \sim 3 \times 10^7$, $\, u \sim 10^4$ cm
s$^{-1}$, $\, l \sim 2.8 \times 10^8$ cm, $\, \rho \sim 5 \times
10^{-4}$ g cm$^{-3} ,$ $\, \eta_{_{T}} \sim 0.9 \times 10^{12} $
cm$^2$ s$^{-1}$; the equipartition mean magnetic field is $B_{\rm
eq} \sim 800$ G and $T \sim 150$. At the bottom of the convective
zone, say at depth $ h_\ast \sim 2 \times 10^{10}$ cm, $\, {\rm
Rm} \sim 2 \times 10^9 ,$ $ \, u \sim 2 \times 10^3 $ cm s$^{-1}$,
$ l \sim 8 \times 10^9$ cm, $ \rho \sim 2 \times 10^{-1}$ g
cm$^{-3} ,$ $\, \eta_{_{T}} \sim 5.3 \times 10^{12} $
cm$^2$s$^{-1}$. Here the equipartition mean magnetic field $B_{\rm
eq} = 3000 $ G and $T \sim 10^7$.  We appreciate that various
estimates for the magnetic Reynolds number and the parameter $T$
for the solar convective zone have been suggested and so we
investigate below the robustness of our results with respect to
$T$. Note also that if we average the parameter $T$ over the depth
of the convective zone, we obtain $T \sim 5$ (see Kleeorin et al.
2003).

\subsection{The turbulent diffusivity}

The simplest order-of-magnitude estimates for magnetic field
turbulent diffusion suggest that it affects all magnetic field
components similarly.  Of course, this does not preclude that a
more detailed parameterization of the turbulent transport
coefficients could result in different estimates for the turbulent
diffusion $\eta_{_{B}}$ of toroidal and $\eta_{_{A}}$ of poloidal
magnetic field components, and Rogachevskii \& Kleeorin (2004)
provide the following estimates for the coefficients $\eta_{_{B}}$
and $\eta_{_{A}}$ for the cases of the weak and strong magnetic
fields (remember that we measure magnetic field strength in units
of the equipartition value $B_{\rm eq}$, and that for the Parker
migratory dynamo the toroidal magnetic field is much stronger than
the poloidal). For the case of weak magnetic field the turbulent
diffusion coefficients are (in units of the reference value
$\eta_{_{T0}}$)

\begin{equation}
\eta_{_{A}} = 1 - {{96} \over 5} \, B^2 \;, \quad \eta_{_{B}} = 1
- 32 B^2, \label{smalB}
\end{equation}
while for strong magnetic fields the scaling is

\begin{equation}
\eta_{_{A}} = {1 \over {8 B^2}}, \quad \quad \eta_{_{B}} = {1
\over {3 \sqrt{2} B}} \; . \label{largeB}
\end{equation}
The transition from one asymptotic form to the other can be
thought of as occurring in the vicinity of $B\sim B_{\rm eq}/4$.

Unsurprisingly, the coefficient of turbulent diffusion of magnetic
helicity $\kappa$ also has a dependence on $B$, namely $\kappa(B)
= 1 - 24 B^2/5$ for weak magnetic field and

\begin{equation}
\kappa(B) = {1 \over 2} \biggl( 1 + {{3 \pi} \over {40 B}} \biggr)
\end{equation}
in the strong field limit. The theory gives more general formulae
for these asymptotical expressions  (see Rogachevskii \& Kleeorin
2004 and Appendix A).

We note that the turbulent diffusion estimates depend on the
details of magnetic field evolution during which the magnetic
helicity accumulated. In particular, the initial ratio between
magnetic and kinetic energy appears in the complete equations of
Rogachevskii \& Kleeorin (2004). We appreciate the importance of
this factor which is almost unaddressed in existing papers in
dynamo theory. However, taking into account the scope of this
paper, we accept (rather arbitrarily) that dynamo action starts in
an (almost) non-magnetized medium. Also, we neglect effects of
possible inhomogeneities in the background turbulence.

\subsection{Nonlinear advection}

Our model contains a inhomogeneous nonlinear suppression of
turbulent magnetic diffusion, which causes turbulent diamagnetic
(or paramagnetic) effects, i.e. a nonlinear advection of magnetic
field which is not the same for the toroidal and poloidal parts of
the magnetic field. The corresponding velocities were calculated
by Rogachevskii \& Kleeorin (2004) yielding

\begin{eqnarray*}
\vec{V}^{A} &=& {32 \over 5} \, B^{2} \, \biggl[ \,
\vec{\Lambda}_B + 3 \vec{\Lambda}_\rho - {{\vec{e}_{r} + \cot
\theta \, \vec{e}_{\theta} }\over {r} }\biggr] \;,
\\
\vec{V}^{B} &=& {32 \over 5} B^2 \biggl[ 3\vec{\Lambda}_\rho-
{\vec{e}_{r} + \cot \theta \, \vec{e}_{\theta} \over {r}}\biggr]
\;
\end{eqnarray*}
for a weak magnetic field, and
\begin{eqnarray*}
\vec{V}^{A} &=& - {1 \over 3 \sqrt 8 B} \biggl[\vec{\Lambda}_B + 2
{\vec{e}_{r} + \cot \theta \, \vec{e}_{\theta} \over {r}} \biggr]
+ {5 \over 16 B^2} \vec{\Lambda}_\rho \;,
\\
\vec{V}^{B} &=& {4 \over 3 \sqrt 8 B} {\vec{e}_{r} + \cot \theta
\, \vec{e}_{\theta} \over {r}} + {5 \over 16 B^2}
\vec{\Lambda}_\rho \;
\end{eqnarray*}
for strong fields. Here $\vec{\Lambda}_B = (\vec{\nabla}
\vec{B}^2) / \vec{B}^2 $, $\vec{e}_r$ and $\vec{e}_\theta$ are
unit vectors in the $r$ and $\theta$ directions of spherical polar
coordinates, $[\vec{\Lambda}_\rho]_r=-d\ln \rho/dr$, and
$[\vec{\Lambda}_B]_r=d\ln B^2/dr$.

\subsection{The rotation law}
\label{rotlaw} In the region $0.7\le r\le 1$ we used an
interpolation on the rotation law derived from helioseismic
inversions. This was extended to include a tachocline region by
interpolating between the helioseismic form at $r=0.7$ and solid
body rotation at $r=r_0$ (see also Moss \& Brooke 2000). Our
choice $r_0=0.64$ gives a rather broad tachocline, but simplifies
the numerics.  Fig.~1 shows contours $\Omega={\rm constant}$.

\section{Results}

\subsection{Numerical implementation}

We simulated the model described above in a meridional cross-section of
a spherical shell with $0 \le \theta \le 180^\circ$ and $0.64
\le r \le 1$.

\begin{figure}
\centering
\includegraphics[width=9cm]{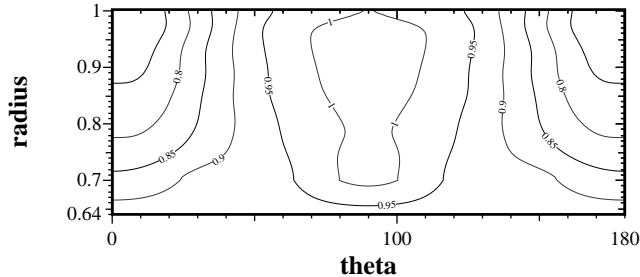}
\caption{\label{Fig1} Isocontours of the rotation law used in the
numerical simulations.}
\end{figure}

The region was divided (rather arbitrarily) into 3 domains, namely
$0.64 \le r < 0.7$, $0.7 \le r \le 0.8$ and $0.8 \le r \le 1$ and
these were identified with the domains of the deep, middle and
shallow active regions of Sect.~2. We attempt to identify the
relative volume occupied by current helicity of `improper' sign
with $N^*/N$.

\subsection{A nonlinear solution}

Our simulations show that the dynamo model leads to a steadily
oscillating magnetic configuration for a quite substantial domain
in the parameter space. These parameters seem acceptable when
compared with current ideas in solar physics. We present here as a
typical model with steady oscillations the case $C_\alpha = -5$,
$\, C_\omega = 6 \times 10^4$ (i.e. $D= - 3 \times 10^5$), $\,
C_1=C_2 = 1$, $\, C_3 = 0.5$, $\, T=5$ and $(2R/l)^2 =300$. (With
this value of $C_\omega$, marginal excitation occurs when
$C_\alpha \approx -4$.) Of course, we are far from understanding
helicity transport inside the Sun well enough  to determine the
numerical value of these parameters. The parameter set chosen
gives a realistic time scale for the cycle period (about 10
years), but with  a rather small nominal value of the turbulent
diffusivity coefficient $\eta_{_{T0}}$, i.e. this is how we choose
to resolve the well-known problem with the length of solar cycle
in the context of mean field dynamo models. The value $|C_\alpha|$
(and $|D|$) chosen is perhaps larger then expected because we use
the profile $\chi^v = \sin ^2 \theta \cos \theta$, which
significantly reduces the mean value of $\chi^v$ over the domain
compared to that with the `standard' $\chi^v= \cos\theta$.

We demonstrated robustness with respect to the value of the
parameter $T$, which is associated with the magnetic Reynolds
number: a uniform increase by two orders of magnitude makes quite
small changes to our results, as does allowing a tenfold increase
from top to bottom of the convection zone. When $T=0.5$ (i.e.
smaller by a factor of 10 than  in the basic run described above)
we still obtain regular oscillations and the magnetic energy
increases by a factor  of 2 or 3 only. However, when $T$ is
significantly  smaller than 0.5,  the solution becomes irregular.
For $T=5$ we also verified that the differences between density
parameter [Eq.~(\ref{rho})] $a=0$ (i.e. uniform density) and
$a=0.3$ were small, and that allowing a radial dependence of
$\chi^v$ also caused only small changes.

\begin{figure}
\centering
\includegraphics[width=9cm]{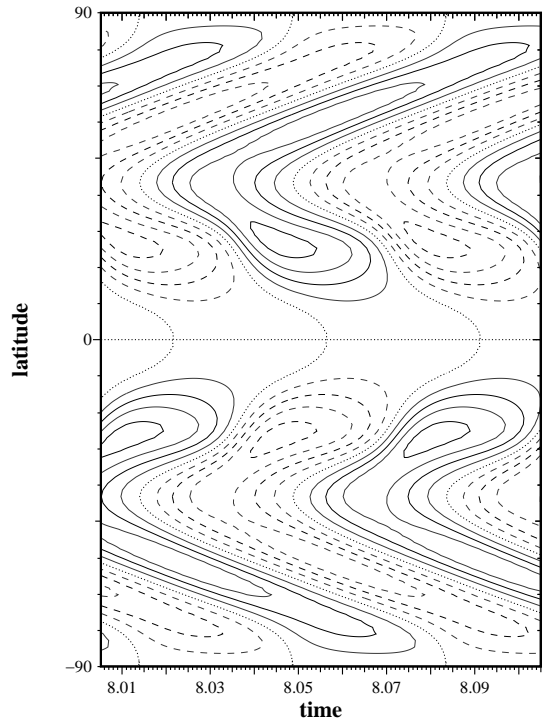}
\caption{\label{Fig2} The near-surface ($r=0.94$) butterfly
diagram of the mean magnetic field. Contours are equally spaced,
solid represent positive values, broken negative, and the zero
contour is shown as dotted.}
\end{figure}

\begin{figure}
\centering
\includegraphics[width=9cm]{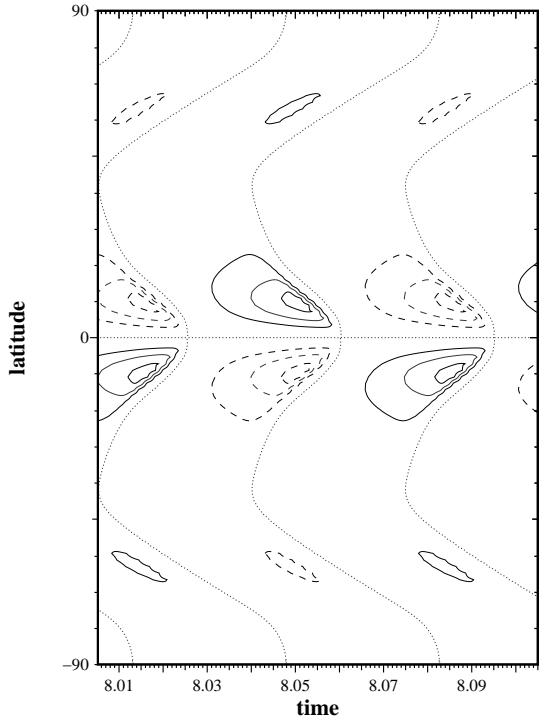}
\caption{\label{Fig3} The butterfly diagram of the mean magnetic
field for the region just above the interface ($r=0.70$).Contours
are equally spaced, solid represent positive values, broken
negative, and the zero contour is shown as dotted.}
\end{figure}

\begin{figure}
\centering
\includegraphics[width=9cm]{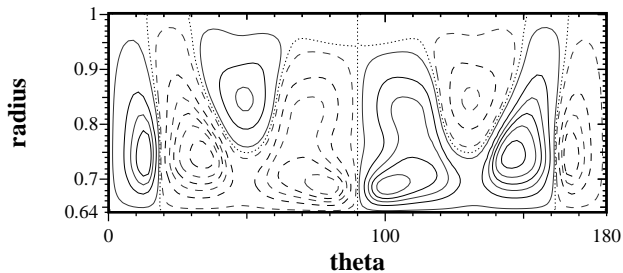}
\caption{\label{Fig4} The toroidal magnetic field distribution at
an instant just after the minimum of magnetic activity. Contours
are equally spaced, solid represent positive values, broken
negative, and the zero contour is shown as dotted.}
\end{figure}

\begin{figure}
\centering
\includegraphics[width=9cm]{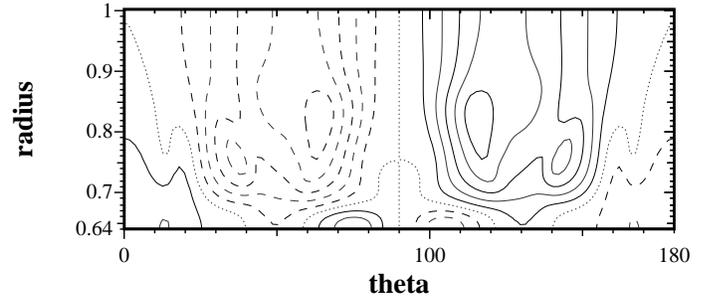}
\caption{\label{Fig5} The current helicity distribution. The
dotted line here indicates the zero level of current helicity.
Contours are equally spaced, solid represent positive values,
broken negative, and the zero contour is shown as dotted.}
\end{figure}

For this typical solution, the magnetic energy $E_m$ measured in the units
of its equipartition value oscillates near the level $E_m \approx
0.12$, and the amplitude of the oscillations is about 0.035. This
means that the averaged magnetic field strength is about 40\% of
the equipartition value. The magnetic configuration can be
described as a system of activity waves which can be presented in
the corresponding butterfly diagrams. In Fig.~2 we show the
near-surface butterfly diagram (at $r=0.94$). Here, a pair of
activity waves migrate from the middle latitudes towards the solar
equator, while another pair migrates from the middle latitudes
towards the poles. We present in Fig.~3 butterfly diagrams for the
region just above the interface (at $r=0.70$). Here, both pairs of
activity waves are much less pronounced in comparison to the
structure shown in Fig.~2. However the equatorward branch now dominates
the poleward. From these synthetic
plots, it seems plausible that the observed butterfly diagram can
be mimicked adequately. The magnetic field structure found in the
simulations is also quite consistent with expectations.

\begin{figure}
\centering
\includegraphics[width=9cm]{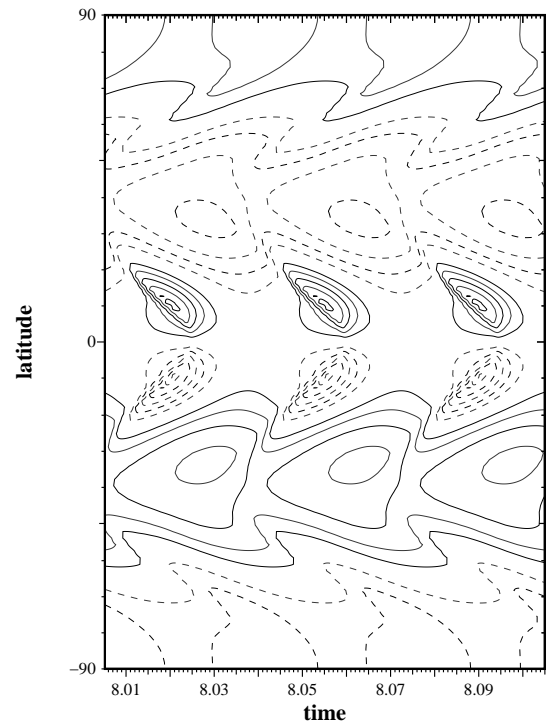}
\caption{\label{Fig6} The butterfly diagram for the current
helicity for the region just above the interface ($r=0.70$).
Contours are equally spaced, solid represent positive values,
broken negative.}
\end{figure}

\begin{figure}
\centering
\includegraphics[width=9cm]{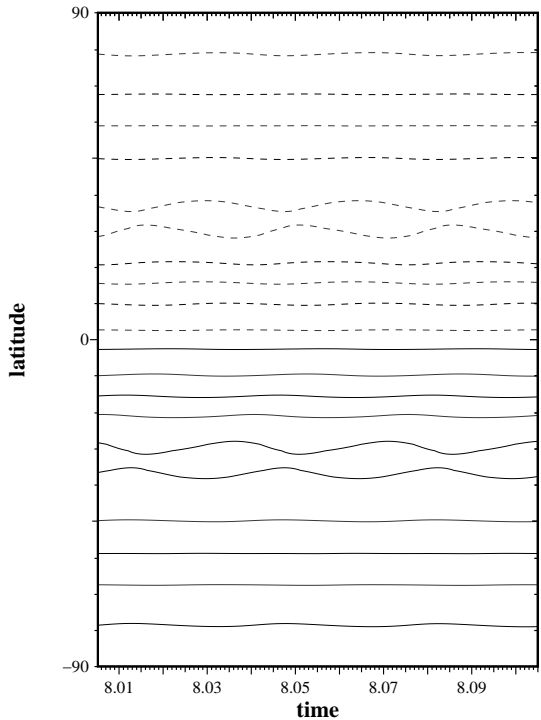}
\caption{\label{Fig7} The near-surface ($r=0.94$) butterfly
diagram for the current helicity. Contours are equally spaced,
solid represent positive values, broken negative.}
\end{figure}

\begin{figure}
\centering
\includegraphics[width=8cm]{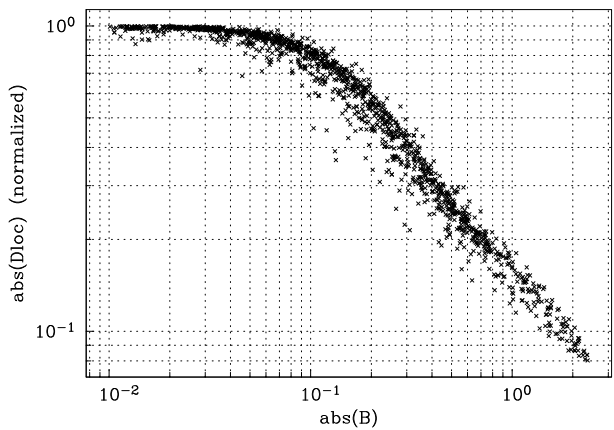}
\caption{\label{Fig8} The normalized local nonlinear dynamo number
at all grid points, as a function of the mean magnetic field.}
\end{figure}

As a typical example, we give in Fig.~4 the toroidal magnetic
field distribution for an instant soon after the minimum of
magnetic energy. The current helicity distribution at the same
time is given in Fig.~5. Here the dotted line indicates the zero
contour of current helicity. The helicity distribution is
antisymmetric with respect to the solar equator, but there are
sign changes inside each hemisphere.  If the helicity is basically
positive in a given hemisphere (e.g. the southern), a region of
negative helicity can be isolated near to the equator at the base
of convective zone. The other region of opposite polarity in the
helicity distribution is located near to the poles. Near the
bottom of the convection zone, the helicity pattern migrates in a
similar way to the toroidal field, and the corresponding butterfly
diagram is given in Fig.~6. Quite unexpectedly, the helicity
pattern near the surface does not demonstrate any pronounced
migration (see Fig.~7).

Of course, we cannot claim a steady oscillating solution obtained
for a rather arbitrary set of parameters should be directly
confronted with proxies of solar activity (see Obridko \& Shelting
2003). We note however that the solution obtained reproduces
remarkably well some features of the solar cycle expected from
dynamo theory and the observational data. Apart from a
conventional equatorward migration, it demonstrates that the
activity cycle is a complicated phenomenon which involves the Sun
as a whole. We see an poleward migration at higher latitudes which
is known from the polar faculae data (Makarov et al. 2001 and
references therein, which give a modern viewpoint of the long-term
research in this area) and from simple illustrative dynamo models
(Kuzanyan \& Sokoloff 1995, 1997). Such a pattern is also seen in
the torsional oscillations, both as observed and as modelled by
Covas, Moss \& Tavakol (2004) -- these are very plausibly
intimately linked to the magnetic field variations. The magnetic
field configuration looks quite simple and smooth for the surface
butterfly diagram of toroidal magnetic field only. We see various
magnetic field reversals inside the Sun. Such reversals have been
suggested by many experts in solar activity whose analysis was not
restricted to sunspot data (e.g. Benevolenskaya et al. 2002). The
dynamo wave at the base of convective zone is much sharper and and
localized (Fig.~3) than that nearer the surface (Fig.~2) -- the
latter appears closer to the current understanding of the solar
cycle. Of course, it is at present unclear just what is the
relation between the sites of field production by the dynamo and
the manifestation of sunspots at the surface. The current helicity
distribution is more complicated at the base of convective zone
compared to that near the solar surface. This  is in general
agreement with the observational information concerning the radial
distribution of solar helicity (Sect.~2).

The normalized local nonlinear dynamo number $D_N = \alpha(B) /
[\eta_{_{A}}(B) \eta_{_{B}}(B)]$ is shown in Fig.~8 at every point of the
computational grid, as a function of mean magnetic field.
Here $\alpha(B)$ is normalized by the local value of
$\alpha(B=0)$.  The nonlinear dynamo number decreases with
increase of the mean magnetic field. The latter dependence implies
the saturation of the growth of the mean magnetic field in the
nonlinear mean field dynamo. Note that the dynamo number which is
based on the hydrodynamic part $\alpha^v$ of the $\alpha$-effect
increases with the mean magnetic field. This shows the very
important role of the magnetic part $\alpha^m$ of the
$\alpha$-effect, which causes the saturation  of the growth of the
mean magnetic field.

\subsection{The helicity distribution}

We need to reduce the numerical data from our modelling to a form
comparable with the observations available. The important point is
that the resolution of  the helicity observations is very
substantially lower than that of the sunspot data,  not to mention
that of the dynamo simulations. The following procedure is applied
to reduce the resolution of the numerical data, and so allow
a meaningful comparison with the observations.

We isolate a region $60^\circ < \theta < 120^\circ$, i.e. a
$60^\circ$--belt centered on the equator, because helicity data
are available for this equatorial domain only. We separate this region into
a deep part, $0.64 \le r \le 0.8$, and a shallow part with
$r>0.8$, and consider one hemisphere only, say the Northern (the
simulated data are strictly antisymmetric with respect to the
solar equator). Let $D_+$ and $D_-$ be the volumes inside each
region where $\chi$ has a positive and negative sign respectively.
We calculate the values $I_+=\int_{T_c} \int_{D_+} \chi^c dV \,
dt$ and $I_-=  \int_{T_c} \int_{D_-} \chi^c \, dV \, dt$, where
$T_c$ is the half length of the activity cycle (note that $I_-$ is
negative).

From our basic run, we obtain the following values of the helicity
integrals. For the `deep' region ($0.64 \le r \le 0.8$), we obtain
$I_- = - 5.4 \times  10^{-5}$ and $I_+ = 2.1 \times 10^{-5}$,
while for the `shallow' region $0.8 \le r \le 1.0$ we obtained
$I_- = - 2.2 \times 10^{-4}$ and $I_+ =0$. The clear difference in
helicity distribution between deep and shallow regions remains
robust when the density parameter $a$ is reduced to $0.3$ [see Eq.
(\ref{rho})]. For the deep region ($0.64 \le r \le 0.8$), we then
obtain $I_+ = - I_- = 4.4 \times 10^{-5}$ (of course, the equality
is a pure coincidence) while for $0.8 \le r \le 1$ we obtained
$I_- = - 2.3 \times 10^{-4}$ and $I_+ =0$.

We conclude that the available observational data concerning the
radial distribution of current helicity seems to be consistent
with the corresponding differences in numerical model. We consider
the observed radial dependence of the current helicity as an
observational manifestation of a structure similar to that
presented in the numerical models.

Note that the choice of the latitudinal and radial  belts in which
the helicity integrals are calculated  does affect significantly the
numbers above. For our basic run, calculating the helicity
integrals for the whole northern hemisphere we obtain $I_- = - 2.7
\times 10^{-3}$ and $I_+ = 7.2 \times 10^{-4}$ for $0.7 \le r \le
0.8$, $I_- = - 5.9 \times 10^{-3}$ and $I_+ = 3.6 \times 10^{-4}$
for $0.8 \le r \le 0.9$ and $I_- = - 6.1 \times 10^{-3}$, $I_+ =
3.5 \times 10^{-4}$ for $0.8 \le r \le 1.0$. Obviously, these
values of helicity integrals calculated for these more arbitrarily
chosen belts are less impressive than the previous, where
the belts were isolated on the basis of snapshots of the helicity
distribution. The important thing is that a link between helicity
integrals and depth is still visible here.

We stress that the available observational data, as well as the
nature of the dynamo model, do not allow any quantitative
description of the radial helicity distribution. The best that we
can hope to do is to isolate some link between these quantities.
The important result is that such a link appears to exist, without
reference to a particular choice of boundaries. We stress this
fact and do not take exactly the same boundaries in for shallow,
middle and deep regions throughout the whole paper.

\section{Discussion}

In this paper we have demonstrated that the available
observational data concerning solar current helicity give some
hints concerning its radial distribution. The active regions
clearly associated with the upper layers of solar convective zone
demonstrate a significantly more homogeneous distribution of
the current helicity than the deeper regions. We interpret this as
an observational indication that the structure of the solar
activity wave deep inside the Sun is substantially more
complicated than near its surface. In contrast to a rather smooth
structure of the surface activity wave with the dominant pattern
propagating from the middle latitudes to the equator, we expect a
more complicated structure of activity waves deep inside the Sun.
In particular, the waves with "wrong" polarity deep inside the Sun
are expected to be more important compared to the main wave than
nearer the surface.

We have demonstrated that the scenario of solar dynamo based on
magnetic helicity conservation arguments can be extended to
include radial dependence. This scenario leads to a steady
oscillatory solution in a substantial domain of the parametric
space, of a form that is at least consistent with our basic
understanding of internal solar structure. If we choose a more
extreme parameter set,  it is natural that we will need to include
more effects (say, buoyancy) into the dynamo saturation
mechanisms.

Slightly unexpectedly, we note that the results of dynamo
simulations are remarkably close to the available magnetic
helicity observations. The structure of dynamo waves deep inside
the convective zone is much sharper and more complicated than the
smooth surface structure. The waves of "wrong" polarity of
helicity are pronounced in deeper layers and almost undetectable at the
surface. We hope that this is an indication that our theoretical
understanding of the solar dynamo has some observational support
from helicity data. Of course, we stress that the very preliminary
nature both of the topic and of our model prevents any strong
conclusion, and that more observational and theoretical efforts
are required to support our inferences. However, in any case the
result obtained is perhaps as good as  could be
expected at the moment.

We emphasize that the ability of the observations to support (or
reject) theoretical ideas concerning the radial properties of the
solar activity wave is highly nontrivial. In contrast, we can
neither support nor reject a scenario suggested by Choudhuri et
al. (2004) who believe that the number of active regions violating
the polarity law should be significantly larger at the beginning
of the cycle rather in the later phases. Some tendency of this
kind is visible in Tables 5 and 6, but the data are insufficient
to support any firm statement. Further studies of a larger sample
of active regions (cf. Bao et al. 2000, 2002) may help to address
this point.

Note that the helicity distribution presented in Fig.~5 could be
represented as a propagation of the activity wave from one
hemisphere to the other. Suppose that a wave of negative helicity
penetrates from northern hemisphere into the southern, where
positive helicity dominates. Such a penetration of an activity
wave into a "wrong" hemisphere was investigated for Parker
migratory dynamo by Galitski et al. (2005). They estimated the
scale of penetration as about a dozen degrees in latitude, which
seems broadly consistent with Fig.~5.

In our basic numerical model, the helicity close to the surface
does not exhibit any migration. This perhaps seems quite
unexpected, but  does not directly contradict any observational or
theoretical knowledge. Note that the butterfly diagrams for the
mean surface poloidal magnetic field exhibit standing, rather
than propagating, waves (Obridko \& Shelting 2003).

\section*{Acknowledgments}

We are indebted to Axel Brandenburg for illuminating discussions.
The research was supported by grants 10233050, 10228307,
10311120115 and 10473016 of National Natural Science Foundation of
China, and TG 2000078401 of National Basic Research Program of
China. A very significant part of this work was performed during
our (NK, DM, IR, DS) visit to the Isaac Newton Institute for
Mathematical Sciences (University of Cambridge). DS and KK are
grateful for support from the Chinese Academy of Sciences and NSFC
towards their visits to Beijing. KK would like to acknowledge
support from RFBR under grants 03-02-16384 and 05-02-16090. DS is
grateful for financial support from INTAS under grant 03-51-5807
and RFBR under grant 04-02-16068. DM acknowledges support from the
Royal Society during a visit to Moscow.

\appendix

\section{Quenching functions}

The quenching functions $\phi_{v}(B)$ and $\phi_{m}(B)$ appearing
in the nonlinear $\alpha$ effect are given by
\begin{eqnarray}
\phi_{v}(B) &=& {1\over 7} [4 \phi_{m} (B) + 3 L(\sqrt{8} B)] \;,
\label{L5} \\
\phi_{m} (B) &=& {3 \over 8 B^2} \biggl[1 - {\arctan (\sqrt{8} B)
\over \sqrt{8} B} \biggr] \; \label{L6}
\end{eqnarray}
(see Rogachevskii \& Kleeorin 2000), where $ L(y) = 1 - 2 y^{2} +
2 y^{4} \ln (1 + y^{-2}) .$

The nonlinear turbulent magnetic diffusion coefficients for the
mean poloidal and toroidal magnetic fields, $ \eta_{_{A}}(B) $ and
$ \eta_{_{B}}(B) $, and the nonlinear drift velocities of poloidal
and toroidal mean magnetic fields, $\vec{V}^{A}(B)$ and
$\vec{V}^{B}(B)$, are given in dimensionless form by
\begin{eqnarray}
\eta_{_{A}}(B) &=& A_{1}(4 B) + A_{2}(4 B) \;,
\label{D1} \\
\eta_{_{B}}(B) &=& A_{1}(4 B) + {3 \over 2} [2 A_{2}(4 B) -
A_{3}(4 B)] \;,
\label{D2}\\
\vec{V}^{A}(B) &=& V_1(B) {\vec{\Lambda}_B \over 2} + {V_2(B)
\over r} \, (\vec{e}_{r} + \cot \theta \, \vec{e}_{\theta}) +
\vec{V}_\rho(B)\;,
\nonumber\\
\label{D3}\\
\vec{V}^{B}(B) &=& {V_3(B) \over r} \, (\vec{e}_{r} + \cot \theta
\, \vec{e}_{\theta}) + \vec{V}_\rho(B) \;, \label{D4}
\end{eqnarray}
where
\begin{eqnarray*}
V_1(B) &=& {3 \over 2} A_{3}(4 B) - 2 A_{2}(4 B) \;,
\\
V_2(B) &=& {1 \over 2} A_{2}(4 B) \;,
\\
V_3(B) &=& {3 \over 2} [ A_{2}(4 B) - A_{3}(4 B)] \;,
\\
\vec{V}_\rho(B) &=& {1 \over 2} \vec{\Lambda}_\rho [-5A_{2}(4 B) +
3 A_{3}(4 B)] ,
\end{eqnarray*}
(Rogachevskii \& Kleeorin 2004).

The functions $ A_{k}(y) $ are
\begin{eqnarray*}
A_{1}(y) &=& {6 \over 5} \biggl[{\arctan y \over y} \, \biggl(1 +
{5 \over 7 y^{2}} \biggr) + {1 \over 14} L(y) - {5 \over 7y^{2}}
\biggr] \;,
\\
A_{2}(y) &=& - {6 \over 5} \biggl[{\arctan y \over y} \, \biggl(1
+ {15 \over 7 y^{2}} \biggr) - {2 \over 7} L(y) - {15 \over
7y^{2}} \biggr] \;,
\\
A_{3}(y) &=& - {2 \over y^{2}} \biggl[ {\arctan y \over y} \,
(y^{2} + 3) - 3 \biggr] \; .
\end{eqnarray*}

The nonlinear quenching of the turbulent magnetic diffusion of the
magnetic helicity is given by
\begin{eqnarray}
\kappa(B) = {1 \over 2} \biggl[1 + A_{1}(4B) + {1 \over 2}
A_{2}(4B) \biggr] \; . \label{D10}
\end{eqnarray}

\end{document}